\documentclass[%
 reprint,
 amsmath,amssymb,superscriptaddress,
 aps,
 prd,
 floatfix
]{revtex4-2}

\usepackage{graphicx}
\usepackage{float}
\usepackage{physics}
\usepackage[dvipsnames]{xcolor}
\usepackage[shortlabels]{enumitem}
\usepackage{bigints}
\usepackage{comment}

\definecolor{change}{rgb}{0., 0., 0.}

\usepackage[normalem]{ulem}

\newcommand{\calO}{{\cal O}}
\newcommand{\calZ}{{\cal Z}}

\newcommand{\hmu}{\, {\hat{\mu}}}

\newcommand{\lat}[2]{{{#1}^3\!\!\times\!\!{#2}}}

\allowdisplaybreaks
\begin{document}

\title{Expansion schemes for finite density QCD equation of state: which one is the best?}

\author{Szabolcs Bors\'anyi}
\affiliation{Department of Physics, Wuppertal University, Gaussstr.  20, D-42119, Wuppertal, Germany}

\author{Zolt\'an Fodor}
\affiliation{Pennsylvania State University, Department of Physics, State College, PA 16801, USA}
\affiliation{Pennsylvania State University, Institute for Computational and Data Sciences, University Park, PA 16802, USA}
\affiliation{Department of Physics, Wuppertal University, Gaussstr.  20, D-42119, Wuppertal, Germany}
\affiliation{Institute  for Theoretical Physics, ELTE E\"otv\"os Lor\' and University, P\'azm\'any P. s\'et\'any 1/A, H-1117 Budapest, Hungary}
\affiliation{J\"ulich Supercomputing Centre, Forschungszentrum J\"ulich, D-52425 J\"ulich, Germany}

\author{Jana N. Guenther}
\affiliation{Department of Physics, Wuppertal University, Gaussstr.  20, D-42119, Wuppertal, Germany}

\author{Piyush Kumar}
\affiliation{Department of Physics, Wuppertal University, Gaussstr.  20, D-42119, Wuppertal, Germany}

\author{Paolo Parotto}
\affiliation{Dipartimento di Fisica, Universit\`a di Torino and INFN Torino, Via P. Giuria 1, I-10125 Torino, Italy}

\author{Attila P\'asztor}
\affiliation{Institute  for Theoretical Physics, ELTE E\"otv\"os Lor\' and University, P\'azm\'any P. s\'et\'any 1/A, H-1117 Budapest, Hungary}
\affiliation{MTA-ELTE Lendület "Momentum" Strongly Interacting Matter Research Group, Budapest, Hungary}

\author{Chik Him Wong}
\affiliation{J\"ulich Supercomputing Centre, Forschungszentrum J\"ulich, D-52425 J\"ulich, Germany}
\affiliation{Department of Physics, Wuppertal University, Gaussstr.  20, D-42119, Wuppertal, Germany}

\date{\today}

\begin{abstract}
The finite density equation of state of QCD is a fundamental quantity for understanding the phase structure of the theory, and also for the realistic modeling of the QCD medium in simulations. While direct lattice simulations at finite density are extremely challenging, several expansion schemes have been devised to extrapolate into the QCD phase diagram. All of these schemes are based on susceptibilities of conserved charges, i.e. derivatives of the pressure with respect to the chemical potential(s). In this work we set out to determine which of these schemes possesses the best convergence properties. To this end, we employ extreme-statistics lattice simulations on a $\lat{16}{8}$ lattice, which allow us to obtain susceptibilities up to order eight with remarkably small errors. We compare in particular two methods, based on contours of constant (normalized) baryon density and of constant entropy density. We find that the latter converges more quickly, effectively requiring susceptibilities only up to order four to cover the chemical potential range up to at least $\mu_B \approx 500$~MeV. We also extend the comparison to the three dimensional space of chemical potentials, and find that the entropy-based method consistently shows superior convergence. 
\end{abstract}

\maketitle

\section{Introduction}

The characterization of the thermodynamic properties of quantum chromodynamics (QCD) is a fundamental goal of high energy nuclear physics, which is the subject of extensive investigations by means of theoretical calculations and large nuclear collision experiments. 
In particular, knowledge of the QCD equation of state in a variety of regimes is necessary both to understand the phase structure of the theory, as well as to provide the thermodynamic information for modeling strongly interacting matter, e.g. to close the set of equations to be solved in fluid dynamic simulations of heavy ion collisions. Recent progress in the mapping of the QCD phase diagram \cite{Borsanyi:2025ttb} 
further emphasizes the need for a solid theoretical knowledge of the thermodynamic properties of the theory.
New predictions on the location of the chiral critical end-point from first principles
functional computations between baryo-chemical potentials $\mu_B$=600 and 700~MeV \cite{Fischer:2026vkc},
as well as the extended density range to be explored by the upcoming CBM experiment
\cite{Collaboration:2026jdg} prompt for precision results in an ever broader range in $\mu_B$.

The most robust tool for calculating QCD thermodynamics are lattice simulations, which in recent years have determined the equation of state at vanishing density to high precision~\cite{Borsanyi:2013bia,HotQCD:2014kol,Borsanyi:2016ksw,Borsanyi:2025dyp}. At finite density, lattice approaches suffer from a complex action problem which drastically increases the cost of direct simulations. Several ways exist to address this shortcoming, either by directly studying the finite-density theory, or by means of extrapolations based on results obtained in the regimes where the complex action problem is absent, namely at zero or purely imaginary chemical potential. The former include Lefschetz thimbles~\cite{Cristoforetti:2012su} or other contour deformations and complex Langevin simulations~\cite{Seiler:2012wz}, but are numerically expensive and cannot yet address QCD phenomenology (see Ref.~\cite{Aarts:2026uiu} for a comprehensive survey). Approaches based on a formulation in the canonical ensemble~\cite{Hasenfratz:1991ax,Alexandru:2005ix,deForcrand:2006ec,Li:2010qf,Danzer:2012vw,Fukuda:2015mva} have also been explored, including a recent study~\cite{Adam:2026rnt} based on the same dataset as the present work, but are limited to smaller volumes. The same limitation applies to the reweighting approach~\cite{Barbour:1997ej}, which consists of writing the path integral in such a way that the complex weight is eliminated by making the observable complex instead. This introduces an overlap problem, caused by the mismatch between the target (desired, e.g. at real $\mu_B$) distribution and the available one (e.g. at $\mu_B=0$), and a sign problem, which causes large cancellations between competing terms and deteriorates the signal-to-noise ratio. To overcome the overlap problem various reweighting schemes have been proposed, e.g. multiparameter reweighting~\cite{Fodor:2001au,Fodor:2001pe,deForcrand:2002pa,Fodor:2004nz} or reweighting based on a sign-quenched ensemble~\cite{Giordano:2020roi,Borsanyi:2021hbk,Borsanyi:2022soo}. These, however, are technically difficult for rooted staggered fermions~\cite{Borsanyi:2023tdp} and are still hindered by the presence of the sign problem.

The second class of methods, on the other hand, has been largely exploited in large-scale investigations, and is based on different recipes to extrapolate toward the finite-density regime. Note that this extrapolation is possible because no singularity is present in the QCD phase diagram at vanishing chemical potential, since the QCD transition is a crossover~\cite{Aoki:2006we}. Most straightforward is the Taylor expansion method~\cite{Gavai:1989ce,Allton:2002zi}, which amounts to calculating derivatives of the quantity of interest -- e.g. the free energy -- with respect to the chemical potential, and constructing the corresponding Maclaurin expansion series. Because the complex action problem is absent for purely imaginary chemical potential, too, an alternative is to calculate the quantity of interest in this regime and analytically continue it to real chemical potential~\cite{Alford:1998sd,deForcrand:2002ci,DElia:2002gd}. These methods have been applied not only to the QCD equation of state, but also to obtain the QCD transition line~\cite{Bellwied:2015rza,Bonati:2018nut,Borsanyi:2020fev,Borsanyi:2024xrx} and other quantities~\cite{Borsanyi:2025kiv} (see e.g.~\cite{Borsanyi:2025ttb} for a recent review).   

For extrapolating the equation of state, both the Taylor and the analytical continuation methods are based on the calculation of derivatives of the free energy, called susceptibilities of conserved charges. However, the numerical cost of these calculations scales with a power of the simulation volume, and higher orders require increasingly large statistics. As an attempt to mitigate this, it was proposed to approximate all higher orders based on a subset of the actual lattice operators, by restricting the computation to specific fixed-order $n$-point function contributions, in a procedure named exponential resummation~\cite{Mondal:2021jxk,Mitra:2022vtf}. While this method is appealing from the technical point of view, numerical tests in Ref.~\cite{Borsanyi:2022soo} have shown no improvement in convergence over the original Taylor expansion.
   
\begin{figure}
    \centering
    \includegraphics[width=\linewidth]{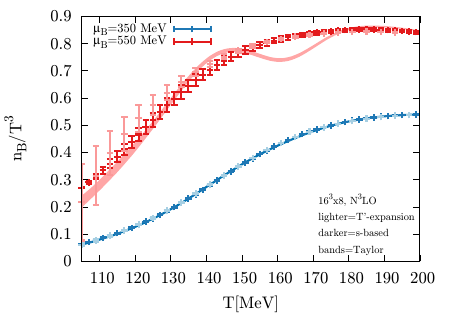}
    \caption{Performance of various extrapolation schemes ($T^\prime$-expansion, $s$-based and Taylor) for the baryon density at finite chemical potential on our $\lat{16}{8}$ lattice.}
    \label{fig:Tprime_s_Taylor}
\end{figure}

In recent years, more sophisticated methods have been developed to construct such extrapolations, which differ by how they combine the available information from zero or imaginary chemical potential and organize it into a systematic expansion. In Ref.~\cite{Borsanyi:2021sxv,Borsanyi:2022qlh} the so-called $T^\prime$-expansion was defined, which amounts to following contours of constant baryon density (normalized by the baryon chemical potential). It was shown that, compared to the Taylor method, larger values of the chemical potentials can be reached with acceptable signal-to-noise ratios (roughly $\mu_B \sim 3.5 \, T$, compared to $\mu_B \sim (2-2.5) \, T$ with Taylor). This is mostly due to the fact that the Taylor expansion is carried out at constant $T$, and hence has to cross over a (pseudo) phase transition as the chemical potential is increased, while the $T^\prime$ expands along paths that follow the crossover line from above or below, never crossing it.
More recently, a similar method based on contours of constant entropy density was proposed in Ref.~\cite{Shah:2024img} to predict the location of the QCD critical point, and later employed to obtain the equation of state in Ref.~\cite{Shah:2026pue} and to provide a rigorous exclusion region for the location of the critical point~\cite{Borsanyi:2025dyp}. Similarly to the $T^\prime$ expansion, this method -- to which we will hereafter refer as $s$-based method -- does not suffer from the shortcomings of the naive Taylor expansion, and shows better convergence properties. Anticipating some of our results, we show in Fig.~\ref{fig:Tprime_s_Taylor} the baryon density at different chemical potentials obtained from the $T^\prime$, $s$-based and Taylor methods, where we see how the latter yields non-physical behavior, while the other two are in excellent agreement.

Since QCD possesses three conserved charges -- baryon number $B$, electric
charge $Q$ and strangeness $S$ -- in general three chemical potentials
$\mu_B,\mu_Q,\mu_S$ can be included. In thermodynamics studies, it is most
common to either consider the simple case $\mu_Q=\mu_S=0$, or the so-called
strangeness-neutral case $n_S=0, n_Q= 0.4 n_B$, which imposes conditions typical
of heavy ion collisions. Here, $n_i$ are the charge densities, and are fixed as
such because the colliding nuclei (Pb at LHC and Au at RHIC) carry no
strangeness and have a neutron-to-nucleon ratio around 0.4. However, more
general conditions are also of interest, for two reasons. First, not only heavy
ion collisions rely on the knowledge of the QCD equation of state: e.g. neutron
stars have $n_S \neq 0$ because of $\beta$ equilibrium. Second, even in heavy
ion collisions, individual cells in the fluid dynamic modeling of the expanding
medium can have large charge asymmetries, and correspondingly nonzero (large)
values of all $\mu_B,\mu_Q,\mu_S$~\cite{Plumberg:2024leb}. Both the Taylor
method~\cite{Monnai:2019hkn,Noronha-Hostler:2019ayj,Monnai:2024pvy} and analytic
continuation 
have been generalized to the case of the full 4-dimensional space of temperature
and chemical potentials, with the latter showing better convergence as in the
case with only a baryon chemical potential. In particular, the $T^\prime$ expansion
has been worked out to NLO~\cite{Abuali:2025tbd} in this 4D space, while the $s$-based method to LO~\cite{Shah:2026mzu}.

Both the Taylor method and its more sophisticated alternatives have been employed as a basis to include critical behavior in the correct universality class~\cite{Parotto:2018pwx,Karthein:2021nxe,Kahangirwe:2024cny,DiClemente:2026hne},
where the position and other non-universal parameters of the end-point are set externally.
This procedure is particularly natural in the $s$-based scheme, which by itself is
capable of predicting such a critical point or critical surface~\cite{Shah:2024img,Shah:2026mzu}.

In this work we set out to determine which method is best for extrapolating the equation of state of QCD to finite density. In particular, the question is which method converges faster to the correct answer, and to this end we will perform a thorough comparison of the $T^\prime$ and $s$-based expansion schemes, since the Taylor approach is evidently inferior in terms of convergence, as shown in Fig.~\ref{fig:Tprime_s_Taylor}. We are in the position to perform this comparison thanks to an extreme-statistics data set we have generated on our $\lat{16}{8}$ lattice, already employed in Refs.~\cite{Adam:2025phc,Adam:2026rnt}, which can count on $\calO ( 3 \cdot 10^5)$ configurations per temperature in the range $T=110-200$~MeV, and in particular $\calO ( (1 - 4) \cdot 10^6)$ in the range $T=115-180$~MeV. This statistics allows us to obtain susceptibilities of up to order eight with great precision (see Ref.~\cite{Adam:2025phc}). This in turn allows us to construct the $T^\prime$ and $s$-based expansions to the same order (NNNLO), and thus to examine if and how convergence of the series is approached. 

In Ref.~\cite{Adam:2026rnt} we presented a new construction to reach finite density while completely avoiding an extrapolation in the chemical potential. Contrary to traditional reweighting, this procedure is not affected by staggered rooting. A shortcoming it still shares with reweighting is the practical limitation to small simulation volumes ($\sim~20-25~\mathrm{fm}^3$). Since Ref.~\cite{Adam:2026rnt} was based on the same high precision data set used here, the direct results obtained there will serve as an ideal testbed to benchmark the expansions we construct in this work.

Since both expansions depend on high order derivatives of simulation data, in order to correctly compute them and estimate the statistical and systematic errors a suitable interpolation scheme must be found. To this end we employ a stochastic Bayesian approach to quintic spline fitting that gives access to higher order derivatives of the susceptibilities.

The paper is organized as follows. In Section~\ref{sec:schemes} we review the $T^\prime$ and $s$-based expansions, highlighting the similarities and differences between them and compared to the naive Taylor expansion. We then present our numerical results for the series convergence in Section~\ref{sec:results}. Because our dataset consists of all diagonal and off-diagonal susceptibilities up to order eight, in Section~\ref{sec:BQS} we generalize our analysis to arbitrary $\mu_B,\mu_Q,\mu_S$, and present results along selected directions. We summarize our conclusions in Section~\ref{sec:conclusion}. We round up the discussion with two technical appendices. In Appendix~\ref{app:schemes} we resolve an ambiguity in the definition of the schemes and show numerical evidence that the choices in the paper (and existing literature) are the best ones. We then describe the stochastic Bayesian interpolation method in Appendix~\ref{app:fit}.

\section{Expansion schemes}\label{sec:schemes}

The naive Taylor expansion for the QCD equation of state is based on the pressure, which is defined at finite chemical potential as:
\begin{equation}
    p(T,\mu_B) = T^4 \sum_{n=0}^{\rm even} c_n (T) \hmu_B^n \, ,
\end{equation}
where $\hmu_B = \mu_B/T$, the sum is restricted to even terms due to charge conjugation symmetry $\calZ (\mu_B) = \calZ (-\mu_B)$, and the expansion coefficients are related to the conserved charges susceptibilities $\chi_n^B(T)$ by:
\begin{equation} \label{eq:chis}
    c_n (T) = \frac{\chi_n^B(T)}{n!} = \frac{1}{n!} \left. \frac{\partial^n (p/T^4)}{\partial \hmu_B^n} \right|_{\mu_B=0} \, \, .
\end{equation}

\begin{figure}
    \centering
    \includegraphics[width=\linewidth]{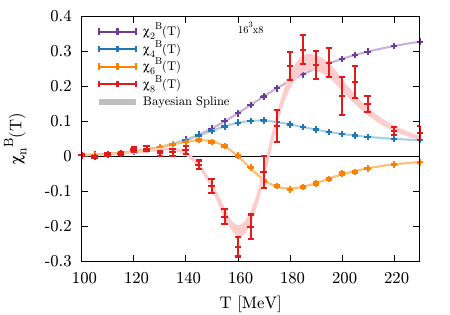}
    \caption{Baryon susceptibilities $\chi_n^B(T)$, which we use as input for both expansion schemes, together with corresponding spline interpolations.}
    \label{fig:chis}
\end{figure}

\begin{figure}
    \centering
    \includegraphics[width=\linewidth]{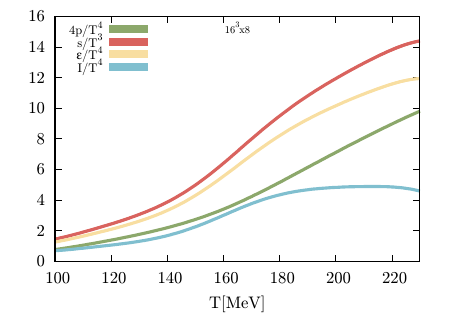}
    \caption{Equation of state -- pressure (green), entropy density (red), energy density (yellow) and trace anomaly (teal) -- at $\mu_B=0$ on our $\lat{16}{8}$ lattice, as functions of the temperature.}
    \label{fig:eosmu0}
\end{figure}

In this work, we use the susceptibilities up to order $\calO(\mu_B^8)$ shown in Fig.~\ref{fig:chis} along with the other thermodynamic observables at $\mu_B=0$ shown in Fig.~\ref{fig:eosmu0}. The multi-dimensional generalizations of these relations are trivial, see e.g. Ref.~\cite{Noronha-Hostler:2019ayj}.

Both the $T^\prime$ and the $s$-based expansions move past the Taylor prescription by avoiding carrying out the extrapolation at constant temperature, and instead following \textit{contours} along which some convenient quantity is fixed. 

\subsection{$T^\prime$ expansion}

Initially developed in Ref.~\cite{Borsanyi:2021hbk}, the defining equation of this expansion scheme is the following:
\begin{equation} \label{eq:Tprime_def}
    \frac{\hat{n}_B (T,\mu_B)}{\hmu_B} = \chi_2^B (T^\prime, 0) \, \, ,
\end{equation}
where we introduced the dimensionless baryon density $\hat{n}_B = n_B/T^3$, and:
\begin{equation} \label{eq:Tprime}
    T^\prime (T,\hmu_B) = T \left( 1 + \sum_{n=2}^{\rm even} \kappa_n(T) \hmu_B^n \right) \, \, .
\end{equation}

Hence, $T^\prime$ is defined as the temperature at which, at $\mu_B=0$, the second baryon susceptibility $\chi_2^B$ takes on the same value as the ratio $n_B (T,\mu_B)/\hmu_B$ does at temperature $T$ and chemical potential $\mu_B$. It was later modified in Ref.~\cite{Borsanyi:2022qlh} to incorporate the approach to the Stefan-Boltzmann limit by redefining the expansion coefficients, but for the purpose of this work it is not essential. 

The $T^\prime$ expansion originated from the observation that the dependence of the baryon density on the (imaginary) chemical potential would satisfy Eqs.~\eqref{eq:Tprime_def},~\eqref{eq:Tprime} to high precision even with only a nonzero coefficient $\kappa_2$ independent of the temperature~\cite{Borsanyi:2021hbk}. This is likely due to the fact that the temperature, chemical potential and quark mass are combined into a single scaling variable at the physical point, which sits in the vicinity of the $O(4)$ chiral critical point. Regardless of its origin, Eqs.~\eqref{eq:Tprime_def},~\eqref{eq:Tprime} define a rigorous expansion scheme which can be exploited to arbitrary order. In Refs.~\cite{Borsanyi:2021hbk,Borsanyi:2022qlh} the coefficients up to order four were determined by means of imaginary $\mu_B$ simulations, but they can be directly related to the susceptibilities $\chi_n^B$ of Eq.~\eqref{eq:chis}, thus requiring only simulations at $\mu_B=0$. By derivating Eq.~\eqref{eq:Tprime_def} wrt $\mu_B$ and evaluating at $\mu_B=0$, one finds up to order $\calO(\hmu_B^8)$:
\begin{align} \label{eq:chistokappas}
    \chi_2^B &= \chi_2^B \, ,\\ \nonumber
    \chi_4^B &= 6 T \kappa_2 \chi_2^{B'} \, ,\\ \nonumber
    \chi_6^B &= 120 T \kappa_4 \chi_2^{B'} + 60 T^2 \kappa_2^2 \chi_2^{B''} \, ,\\ \nonumber
    \chi_8^B &= 5040 T \kappa_6 \chi_2^{B'} + 5040 T^2 \kappa_2 \kappa_4 \chi_2^{B''} \\ \nonumber
    &\,\,\,\,\quad\quad\quad\quad\quad\quad+ 840 T^3 \kappa_2^3 \chi_2^{B'''} \, ,
\end{align}
where the temperature dependence of the $\chi_n^B$ and $\kappa_n$ is understood, and the symbol $^\prime$ indicates derivatives wrt the temperature. In general, an expansion up to $\kappa_n$ requires and includes susceptibilities up to $\chi_{n+2}^B$ and temperature derivatives up to order $n/2$. This expansion, being based on an observable which is already a derivative of the free energy wrt $\mu_B$, automatically starts at the next order, i.e. the next-to-leading order (NLO), which includes up to $\chi_4^B$. 

In this work we employ susceptibilities up to order eight, and determine the coefficients up to $\kappa_6(T)$ by means of the above relations. Once the expansion coefficients $\kappa_n(T)$ are determined, Eq.~\eqref{eq:Tprime_def} is exploited to determine the baryon density $n_B$ at real chemical potential, and from it the pressure and other thermodynamic variables:
\begin{align}
    \hat{p} (T,\hmu_B) &= \hat{p} (T,0) + \bigintssss_0^{\hmu_B} \hat{n}_B (T,\hmu_B^\prime) \, d \!\hmu_B^\prime \, , \\ \nonumber
    \hat{s} (T,\hmu_B) &= 4 \hat{p} (T,\hmu_B) + T \frac{\partial \hat{p}(T,\hmu_B)}{\partial T} - \hmu_B \hat{n}_B(T,\hmu_B) \, , \\ \nonumber
    \hat{\epsilon} (T,\hmu_B) &= \hat{s} (T,\hmu_B) - \hat{p}(T,\hmu_B) + \hmu_B \hat{n}_B(T,\hmu_B) \, ,
\end{align}
where $\hat{p} = p/T^4$, $\hat{s} = s/T^3$, $\hat{\epsilon} = \epsilon/T^4$ are the dimensionless pressure, entropy density and energy density, respectively, and the baryon density from Eq.~\eqref{eq:Tprime_def} is used. The first term in the expression for the pressure provides the input on the zero-density equation of state in the form of an integration constant.

\subsection{$s$-based expansion}

The $s$-based expansion was introduced in Ref.~\cite{Shah:2024img} to estimate the location of the critical point, based on the observation that, if two contours of constant entropy overlap, this signals the onset of a first order transition where the entropy itself is multi-valued.

In this case, the defining equation is:
\begin{equation} \label{eq:sbased_def}
    s(T,\mu_B) = s(T_0,0) \, \, , 
\end{equation}
where:
\begin{equation} \label{eq:sbased_Ts}
    T(T_0,\mu_B) = T_0 + \sum_{n=2}^{\rm even} \frac{\alpha_n (T_0)}{n!}  \mu_B^n \, \, .
\end{equation}

\begin{figure*}
    \centering
    \includegraphics[width=0.49\linewidth]{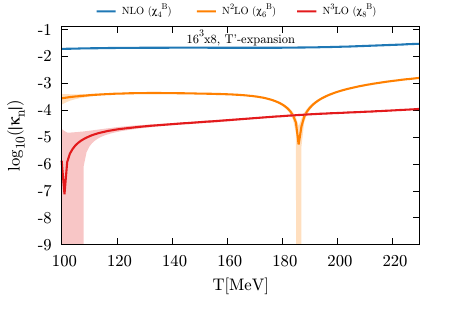}
        \includegraphics[width=0.49\linewidth]{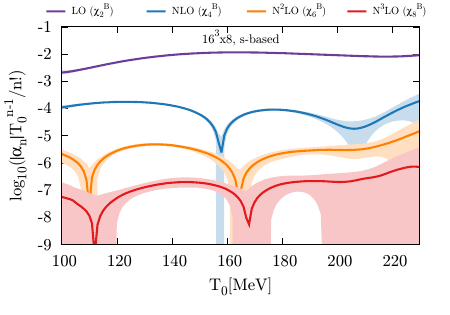}
    \caption{Absolute value of the expansion coefficients for the $T^\prime$ (left) and $s$-based (right) expansions, as functions of the temperature. In the right panel, the coefficients are made dimensionless by multiplying with powers of the temperature.}
    \label{fig:coefficients}
\end{figure*}

Here $T_0$ is defined as the temperature at which, at $\mu_B=0$, the entropy density has the same value it has at finite $\mu_B$ and temperature $T$. Note that, similarly to its original formulation, in Eq.~\eqref{eq:sbased_def} we \textit{do not} use the dimensionless entropy density, nor the dimensionless chemical potential. This means that the expansion coefficients are dimensionful, and in order to properly compare them to one another, they should first be made dimensionless by multiplying with suitable powers of the temperature, as we do in Fig.~\ref{fig:coefficients} in the following. Nothing prevents one to define the extrapolation with different choices, but it was found -- also in the preparation of this work -- that this one yields better convergence properties. On the other hand, the opposite can be said for the $T^\prime$ expansion, which was defined -- and works better -- based on dimensionless quantities as in Eq.~\eqref{eq:Tprime_def}. The coefficients $\alpha_n(T_0)$ can also be directly related to the susceptibilities. Starting from the expansion of Eq.~\eqref{eq:sbased_Ts}, we expand both sides of Eq.~\eqref{eq:sbased_def} and match them order by order up to $\calO(\mu_B^n)$ to obtain $\alpha_n$, as follows. 

The dimensionless entropy at finite chemical potential is defined as:
\begin{equation}
   \hat{s}(T,\mu_B) = \hat{s}(T,0) + 4 \Delta \hat{p} + T \left. \frac{\partial \Delta \hat{p}}{\partial T} \right|_{\hmu_B} - \hmu_B \hat{n}_B \, ,
\end{equation}
where:
\begin{align}
    \Delta \hat{p}(T, \mu_B) &= \sum_{n=2}^{\rm even} \frac{\chi_n^B(T)}{n!} \hmu_B^n \, , \\ \nonumber
    \hat{n}_B(T, \mu_B) &= \sum_{n=2}^{\rm even} \frac{\chi_{n-1}^B(T)}{(n-1)!}\hmu_B^{n-1} \, \, .
\end{align}
With these expressions one can write the entropy density itself as an expansion in $\hmu_B$ as:
\begin{equation}
    \hat{s}(T,\mu_B) = \sum_{n=0}^{\rm even} \hat{s}_n (T) \hmu_B^n \, , 
\end{equation}
with
\begin{equation}
\hat{s}_{n} (T) = \frac{(4-n) \chi^B_{n} (T) + T \chi_{n}^{B'} (T)}{n!} \,\,.
\end{equation}

The left hand side of Eq.~\eqref{eq:sbased_def} written as an expansion reads:
\begin{equation}
    s(T,\mu_B) = T^3 \hat{s}(T,\mu_B) = \sum_{n=0}^{\rm even} T^{3-n} \hat{s}_n(T) \mu_B^n \, \, ,
\end{equation}
where the expansion coefficients of the (dimensionful) entropy density thus read:
\begin{equation}
    \mathcal{S}_{n}(T) = T^{3-n} \hat{s}_{n} (T) = \frac{1}{n!} \frac{d}{dT} \left( T^{4-n} \chi_{n}^{B}(T) \right) \, ,
\end{equation}
and $\mathcal{S}_{0}(T) = s_0(T)$ is simply the (dimensionful) entropy density at $\mu_B=0$. 

Using the expansions for $T$ and $\hat{s}_{n}$ around $T_0$, we can collect terms for various powers of $\mu_B$. Note that the rhs of Eq.~\eqref{eq:sbased_def} only contains the $\mu_B^{0}$ term, which is matched by the $\mu_B^0$ term of the lhs. Solving the contour equation for higher orders thus means imposing that the sum of the terms in the lhs is zero. This yields the $\alpha_{n}(T_0)$ as functions of the starting temperature $T_0$. In the following we suppress $T_0$ as it is clear from the context that all the functions and the derivatives are evaluated at $T_0$. The $\alpha_{n}$ are then given by:
\begin{align}
\alpha_2 &= -\frac{2 \mathcal{S}_2}{\mathcal{S}'_0} \, \, ,\\ \nonumber
\alpha_4 &= - \frac{24}{\mathcal{S}'_0} \left(\mathcal{S}_4 + \frac{\alpha_2}{2} \mathcal{S}'_2 + \frac{\alpha_2^2}{8} \mathcal{S}''_0\right) \, \, ,\\ \nonumber
\alpha_6 &= - \frac{720}{\mathcal{S}'_0} \left( \mathcal{S}_6 + \frac{\alpha_2}{2} \mathcal{S}'_4 + \frac{\alpha_4}{24} \mathcal{S}'_2 + \frac{\alpha_2^2}{8} \mathcal{S}''_2 \right. \\ \nonumber
&\qquad\qquad\qquad\qquad+ \left. \frac{\alpha_2 \alpha_4}{48} \mathcal{S}''_0 + \frac{\alpha_2^3}{48} \mathcal{S}'''_0 \right) \, \, ,\nonumber
\\
\alpha_8 &= -\frac{40320}{\mathcal{S}_0'} \left (\mathcal{S}_8 + \frac{\alpha_2}{2} \mathcal{S}_6' + \frac{\alpha_2^2}{8} \mathcal{S}_4'' + \frac{\alpha_4}{24} \mathcal{S}_4' 
\right.
\nonumber\\
&  + \frac{\alpha_2^3}{48} \mathcal{S}_2'''+ \frac{\alpha_2 \alpha_4}{48} \mathcal{S}_2'' + \frac{\alpha_6}{720}\mathcal{S}_2' \nonumber\\
& \left. + \frac{\alpha_2^4}{384} \mathcal{S}_0^{(4)} + \frac{\alpha_2^2 \alpha_4}{192} \mathcal{S}_0''' + \frac{\alpha_4^2}{1152} \mathcal{S}_0''+ \frac{\alpha_2 \alpha_6}{1440} \mathcal{S}_0'' \right)\,\,,\nonumber
\end{align}
which e.g. for order two reads:
\begin{equation}
        \alpha_2 = - \frac{(T^2 \chi_2^B)'}{s_0'} = - \frac{2T \chi_2^B + T^2 \chi_2^{B'}}{s_0'} \, \, .
\end{equation}

This is the method used in Ref.~\cite{Shah:2024img} to construct the contours of constant entropy density, which the authors employ to order $\alpha_2$, i.e. including susceptibilities up to order two. 

In this work we employ susceptibilities up to order eight, and determine the coefficients up to $\alpha_8(T_0)$ by means of the the direct relations between the $\alpha_n$ coefficients and the susceptibilities, which are: 
\begin{align}
\frac{T^2 \chi_2^B}{2} &= - \int \left(\frac{\alpha_2 \mathcal{S}'_0}{2}\right)dT + C_2 \,,\\ \nonumber
\frac{\chi_4^B}{24} &= \left( \frac{\alpha_2^2}{8} \mathcal{S}_0' \right) - \int \left(\frac{\alpha_4}{24} \mathcal{S}_0' \right)dT + C_4\,,\\ \nonumber
\frac{\chi_6^B}{720 T^2} &= \left( \frac{\alpha_2 \alpha_4}{48}\mathcal{S}_0' - \frac{\alpha_2^2 \alpha_2'}{16} \mathcal{S}_0' - \frac{\alpha_2^3}{48} \mathcal{S}_0'' \right) \\ \nonumber
&\qquad- \int \left(\frac{\alpha_6}{720}\mathcal{S}_0'\right) dT + C_6\,,\\ \nonumber
\frac{\chi_8^B}{40320 T^4} &= \left( \frac{\alpha_2^4}{384} \mathcal{S}_0''' + \frac{\alpha_2^3 \alpha_2''}{96} \mathcal{S}_0' + \frac{\alpha_2^3 \alpha_2'}{48} \mathcal{S}_0'' \right.\\ \nonumber
&\qquad- \left.
\frac{\alpha_2^2 \alpha_4}{192}\mathcal{S}_0'' + \frac{\alpha_2^2 (\alpha_2')^2}{32} \mathcal{S}_0' - \frac{\alpha_2^2 \alpha_4'}{192} \mathcal{S}_0' \right. \\ \nonumber
&\qquad- \left. \frac{\alpha_2 \alpha_4 \alpha_2'}{96}\mathcal{S}_0' + \frac{\alpha_2 \alpha_6}{1440} \mathcal{S}_0' + \frac{\alpha_4^2}{1152}\mathcal{S}_0' \right) \\ \nonumber
&\qquad- \int \left(\frac{\alpha_8}{40320}\mathcal{S}_0'\right) dT + C_8\,,
\label{eq:chistoalphas}
\end{align}
where the integration constants $C_{2-8}$ are also fit parameters. Once the coefficients are determined, the thermodynamics at finite chemical potential can be easily reconstructed as follows.

The entropy at finite temperature and chemical potential is given by Eq.~\eqref{eq:sbased_def}, with $T_0(T,\mu_B)$ implicitly defined through Eq.~\eqref{eq:sbased_Ts}. The pressure then reads:
\begin{align}
    p(T,\mu_B) - p(T_{\rm ref}, \mu_B) &= \int_{T_{\rm ref}}^T s(T',\mu_B) dT'\\ \nonumber
    &= \int_{T_{\rm ref}}^T s_0(T_0(T',\mu_B)) ~dT' \, \, ,
\end{align}
where we had to introduce the integration constant $p(T_{\rm ref}, \mu_B)$ at an arbitrary temperature $T_{\rm ref}$, to be fixed separately. Through the change of variables $T' \to T_0$, the integral becomes:
\begin{equation}
    p(T,\mu_B) - p(T_{\rm ref}, \mu_B) = \int_{T_0(T_{\rm ref},\mu_B)}^{T_0(T,\mu_B)} s_0(T_0) \left. \frac{dT'}{dT_0}\right|_{\mu_B} dT_0 \, \, ,
\end{equation}
where $dT'/dT_0\vert_{\mu_B}$ is determined via Eq.~\eqref{eq:sbased_Ts}.
The baryon density $n_B(T,\mu_B)$ then follows:
\begin{align}
    n_B(T,\mu_B) - n_B(T_{\rm ref}, \mu_B) &= \left. \frac{\partial p(T, \mu_B)}{\partial \mu_B} \right \vert_T \\ \nonumber 
    &=\int_{T_{\rm ref}}^T \left. \frac{\partial s_0(T_0(T',\mu_B))}{\partial \mu_B} \right \vert_{T'} dT' \, \, ,
\end{align}
where the integrand evaluates to
\begin{equation}
    \left. \frac{\partial s_0(T_0(T',\mu_B))}{\partial \mu_B} \right \vert_{T'} = s_0'(T_0(T',\mu_B)) \left. \frac{dT_0}{d\mu_B} \right|_{T^\prime} \, \, ,
\end{equation}
and $dT_0/d\mu_B$ at fixed $T'$ is given by implicit differentiation:
\begin{equation}
    \frac{dT_0}{d\mu_B} = - \frac{\partial T'(T_0,\mu_B)/\partial \mu_B}{\partial T'(T_0,\mu_B)/ \partial T_0} \, \, .
\end{equation}

Since the temperature integral is performed at constant $\mu_B$, we can repeat the change of variables $T' \to T_0$ and obtain the simplified expression:
\begin{equation}
    n_B (T, \mu_B) - n_B(T_{\rm ref}, \mu_B) = -\!\!\int \limits_{T_0(T_{\rm ref},\mu_B)}^{T_0(T,\mu_B)} \!\!s_0'(T_0) \left.\frac{\partial T'}{\partial \mu_B}\right|_{T_0} dT_0 \, \, .
\end{equation}

Once the entropy, pressure and baryon density are known, the energy density is given by:
\begin{equation}
    \epsilon (T,\mu_B) = T s(T,\mu_B) - p(T,\mu_B) + \mu_B n_B(T, \mu_B) \, \, .
\end{equation}

This concludes the description of the two expansions, completely defined by their respective ``master'' equations Eqs.~\eqref{eq:Tprime_def}-\eqref{eq:Tprime} and~\eqref{eq:sbased_def}-\eqref{eq:sbased_Ts}, respectively. All expansion coefficients follow from the same set of  susceptibilities $\chi^B_n$ and the equation of state at $\mu_B=0$ (see
Figs.~\ref{fig:chis}, \ref{fig:eosmu0}).

We have seen that both methods amount to a reorganization of the Taylor coefficients with the goal of improving the convergence of the series. It is obvious that, for high enough order, both should agree among themselves and with the Taylor expansion, as the input information is the same. However, for large scale lattice simulations, it is imperative that the expansion converges quickly, since each subsequent order to be calculated requires a substantial increase in the numerical cost.

The differences between the two methods lie in the specific way the contours are built and followed. The first obvious difference between the two schemes is that they are based on contours of different observables. Second, while both methods define contours by means of an \textit{effective} temperature, they differ by which temperature is written as an expansion. In the $T^\prime$ expansion, it is the temperature defined at $\mu_B=0$, while in the $s$-based method it is the finite-$\mu_B$ temperature which is expanded. One can refer to the first approach as an \textit{explicit} method, and to the second as \textit{implicit}. In fact, it is not obvious that one choice performs better than the other, and we constructed both the $T^\prime$ expansion in the implicit formulation and the $s$-based expansion in the explicit one to compare the respective performance. We compare the convergence properties of both expansions with these alternative choices in Appendix~\ref{app:schemes}. There, we show that the $s$-based \textit{implicit} method we show in the main text outperforms all other options in term of convergence. In general, we observe that \textit{implicit} schemes converge faster, albeit their implementation is more involved. Note that we are in the position to draw these conclusions only thanks to our extreme-statistics dataset, which allows us to see convergence of all methods at high enough order (NNNLO), and additionally to compare to the directs results from Ref.~\cite{Adam:2026rnt}.

\begin{figure*}
    \centering
    \includegraphics[width=0.49\linewidth]{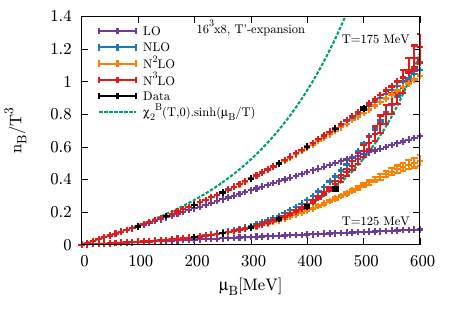}
        \includegraphics[width=0.49\linewidth]{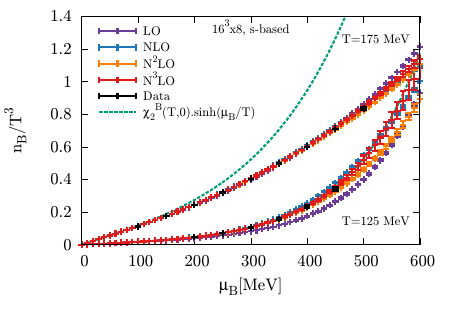}
    \caption{Baryon density as a function of the chemical potential at constant temperatures $T=125,175$~MeV from the $T^\prime$ (left) and $s$-based (right) expansions. Different colors correspond to different orders in the expansion, including up to $\chi_2^B$ (purple), $\chi_4^B$ (blue), $\chi_6^B$ (orange) or $\chi_8^B$ (red). Black points are obtained through the direct method of Ref.~\cite{Adam:2026rnt}. The green dashed line is the result of the hadronic ansatz we employ at $T_{\rm ref}=105$~MeV to fix the integration constant in the $s$-based method.}
    \label{fig:nb_Tconst}
\end{figure*}

\section{Results}\label{sec:results}

In this work we employ our results for conserved charges susceptibilities on a $\lat{16}{8}$ lattice, and construct the $T^\prime$ and $s$-based expansions with the methods detailed above. We show in Fig.~\ref{fig:chis} the baryon susceptibilities up to order $\calO(\mu_B^8)$, together with a spline interpolation (see Appendix~\ref{app:fit} for details). We also determined the equation of state at $\mu_B=0$ on the same lattice to fix the thermodynamics at finite chemical potential, which we show in Fig.~\ref{fig:eosmu0}. In order to estimate the systematic errors stemming from the necessary interpolations, we employ a global fit procedure that is carried out only once for each method. In particular, the coefficients $\kappa_n(T)$ and $\alpha_n(T_0)$, respectively, are modeled by means of basis splines of order five (because derivatives of order three need to be taken). The parameters of the splines are set by minimizing the combined chi square of all input susceptibilities together.
This means that the chi squared function is constructed from the squared deviations of e.g. Eqs.~(\ref{eq:chistokappas}) or Eqs.~(\ref{eq:chistoalphas}) from the data points for $\chi^B_2,\dots,\chi^B_8$ and the trace anomaly at $\mu=0$. We work with a pool of 100 choices for the number and location of the spline node points that define the coefficient functions, generated through a Metropolis-like method. The systematic error is computed
from the variation of the final thermodynamics results (e.g. baryon density, entropy density) depending on the choice of node points. More details on the fit procedure and the analysis of systematic uncertainties are reported in Appendix~\ref{app:fit}.

The first comparison one can carry out is between the expansion coefficients. In order to achieve a quick convergence, subsequent orders should decay quickly, more so than the Taylor coefficients do. In Fig.~\ref{fig:coefficients} we compare the $\kappa_n(T)$ and $\alpha_n(T_0)$ as functions of the temperature, and notice that indeed in both cases the first coefficient is by far the largest. 

In the following we fix the notation as follows. For both schemes, we refer to expansions up to $\calO(\mu_B^2)$ as LO, $\calO(\mu_B^4)$ as NLO,  
$\calO(\mu_B^6)$ as NNLO and $\calO(\mu_B^8)$ as NNNLO (the same will apply to the 3D case in Section~\ref{sec:BQS}). Whenever we compare results at different orders, we stick to the color scheme of Fig.~\ref{fig:chis} to indicate the highest order susceptibility included. 

\begin{figure*}
    \centering
    \includegraphics[width=0.49\linewidth]{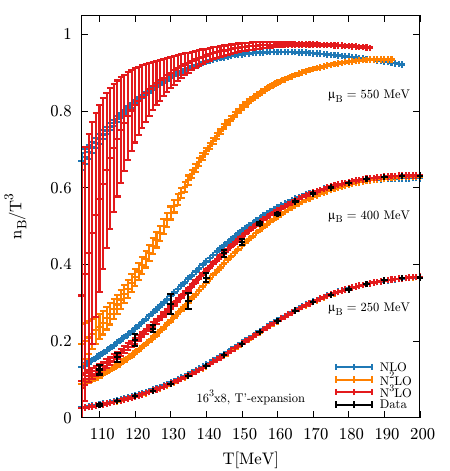}
        \includegraphics[width=0.49\linewidth]{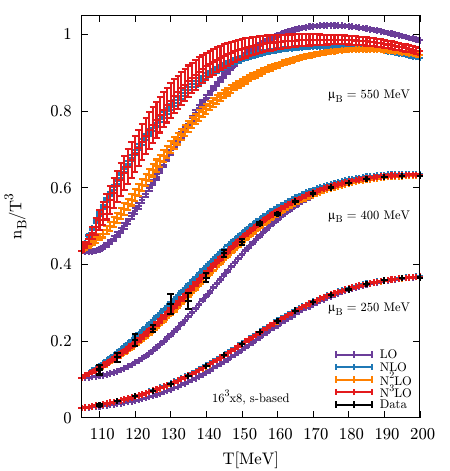}
    \caption{Baryon density as a function of the temperature at constant chemical potentials from the $T^\prime$ (left) and $s$-based (right) expansions. Different colors correspond to different orders in the expansion, including up to $\chi_4^B$ (blue), $\chi_6^B$ (orange) or $\chi_8^B$ (red). Black points are obtained through the direct method of Ref.~\cite{Adam:2026rnt}.}
    \label{fig:nb_muconst}
\end{figure*}

We can now compare results for thermodynamic variables between the two methods and different orders. We focus on the baryon density because it can show the largest deviations, since it does not contain a zero-density term in common to both methods and all orders. We fix the integration constant needed for the $s$-based method by imposing $\hat{p}(T_{\rm ref},\mu_B) = \hat{p}_0(T_{\rm ref}) +  \hat{p}_1(T_{\rm ref}) \cosh(\mu_B/T_{\rm ref})$, hence $n_B(T_{\rm ref}, \mu_B) = T_{\rm ref}^3 \hat{p}_1 (T_{\rm ref}) \sinh(\mu_B/T_{\rm ref})$, with $\hat{p}_1(T_{\rm ref}) = \chi_2^B(T_{\rm ref})$ and $\hat{p}_0(T_{\rm ref})$ obtained from the $\mu_B=0$ equation of state. This choice for the $\mu_B$-dependence of the equation of state is imposed at all orders at $T_{\rm ref}=105$~MeV, thus forcing the agreement at this temperature. Although this choice may appear arbitrary, the $\cosh$ model (based on a fugacity expansion in terms of fixed-baryon number sectors) is extremely accurate at this temperature value. We could check that this is the case at all imaginary $\mu_B$ values we have access to, with several simulation volumes and lattice spacings. This model is motivated by the hadronic nature of the active degrees of freedom, and it is trustworthy as long as $T_{\rm ref}$ is sufficiently below the $\mu_B$-dependent QCD transition temperature $T_c(\mu_B)$.

The large separation of scales between the first and subsequent orders' coefficients in both expansions does indeed reflect into a good convergence at finite chemical potential. In Fig.~\ref{fig:nb_Tconst}, for two constant temperature values $T=125,175$~MeV, we compare the baryon density as a function of the chemical potential at different orders, from the $T^\prime$ (left) and $s$-based (right) expansions. Note that the same color applies to the same order in $\mu_B$, although the $T^\prime$ expansion starts one order ahead due to its formulation. At the larger temperature, all orders agree up to $\mu_B \simeq 500$~MeV, while at the lower one the disagreement appears around $\mu_B=350$~MeV with both schemes. One can see that the $s$-based method seems to converge better, although the difference is not dramatic. Notably, in both methods the $\calO (\mu_B^4)$ expansion agrees with the $\calO (\mu_B^8)$ better than the $\calO (\mu_B^6)$. In Fig.~\ref{fig:nb_muconst} this can be seen in more detail. The baryon density is shown as a function of the temperature at fixed chemical potentials. Here it is evident that the $s$-based approach converges more quickly to the $\calO (\mu_B^8)$ result. We must note that the fact that all orders in the $s$-based expansion are forced to agree at $T_{\rm ref}=105$~MeV helps the convergence of the series. However, the fact the agreement persists to a good approximation at all temperatures indicates that our choice for the integration constant $p(T_{\rm ref}, \mu_B)$ is justified. 
Of course, the choice $T_{ref}=105$~MeV is arbitrary. However, one expects that e.g. at and below $T=125$~MeV matter is still predominantly hadronic. As a consistency check, we added to both panels of Fig.~\ref{fig:nb_Tconst} dashed lines representing the result given by our ansatz for $\hat{p}(T_{\rm ref},\mu_B)$ at the corresponding temperature. This is simply the function $\chi^B_2(T,0) \cdot \sinh(\mu_B/T)$. Its excellent agreement with the $s$-based extrapolation at this temperature, 20 MeV above the reference, is not trivial (see the change in $n_B/T^3$ between 125 and 105~MeV in Fig.~\ref{fig:nb_muconst}) and guarantees the validity of our ansatz. 
In future works reaching larger values of $\mu_B$, criticality must be accounted for. This is readily possible in the $s$-based approach \cite{Shah:2024img}, independently of any model assumption on $\hat{p}(T_{\rm ref},\mu_B)$. Note that $\hat{p}(T_{\rm ref},\mu_B)$ is needed only to compute thermodynamic observables \textit{after} the contours have been determined.
Should a critical point or 1st order region be identified in the future, $T_{\rm ref}$ must be lowered accordingly. At present the possibility to insert physics input at the lowest temperature should be seen as an advantage of the method, which accurately determines the $T$-dependence of the equation of state, and ``imports'' the $\mu_B$ dependence determined by other means. Especially at lower temperature, the disagreement between orders in the $T^\prime$ expansion is much more pronounced.

In Figs.~\ref{fig:nb_Tconst} and \ref{fig:nb_muconst} the black data points indicate the full grand canonical result on this lattice obtained with the method of Ref.~\cite{Adam:2026rnt}. This method yields the would-be infinite-order summation of the Taylor series through a saddle-point representation of the Legendre transform between the grand-canonical and canonical ensembles. The canonical input for the finite-$\mu_B$ results was obtained by reweighting the very same ensembles we use in this work, in the form of $\chi$ coefficients.
Since the results with the method of Ref.~\cite{Adam:2026rnt} do not depend on any extrapolation in $\mu_B$, we label them as ``data''. The accessible range is limited to 450~MeV for this volume due to the residual sign and overlap problems inherent to direct finite-$\mu_B$ approaches. Note that that, as every reweighting, the method of Ref.~\cite{Adam:2026rnt} is limited to small volumes, while the $T'$ and $s$-based resummations are not.

As far as these data points are available, the agreement with both N$^3$LO expansions is excellent.

\begin{figure*}
    \centering
    \includegraphics[width=0.49\linewidth]{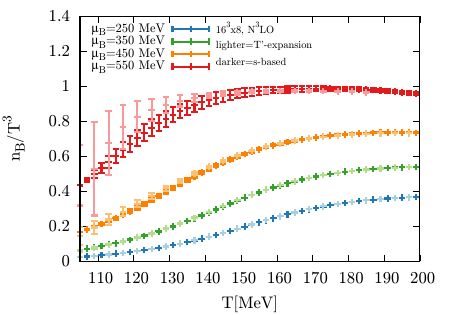}
        \includegraphics[width=0.49\linewidth]{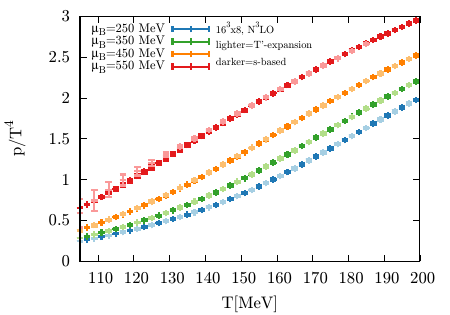} \\
            \includegraphics[width=0.49\linewidth]{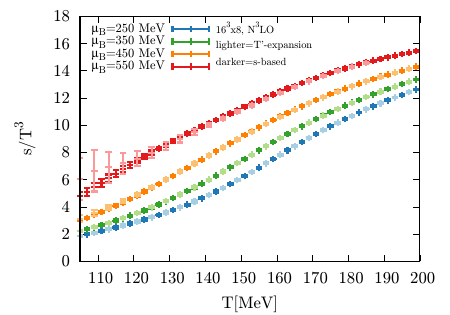}
        \includegraphics[width=0.49\linewidth]{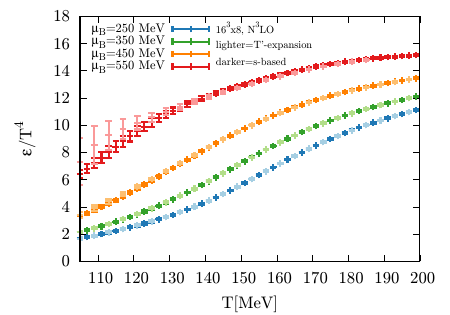}
    \caption{Baryon density, pressure, entropy density and energy density as functions of the temperature at constant chemical potentials from the $T^\prime$ (lighter shades) and $s$-based (darker shades) expansions, at order $\calO (\mu_B^8)$.}
    \label{fig:all_comparison_O8}
\end{figure*}

As we already noted, for high enough order all expansions should agree. We see that this is indeed the case in Fig.~\ref{fig:all_comparison_O8}, where the baryon density, pressure, entropy density and energy density are shown for different $\mu_B$ as functions of the temperature in both methods at the highest order $\calO (\mu_B^8)$. In all cases the agreement is excellent, although the errors in the $s$-based expansion are smaller at lower $T$ due to the condition on $p(T_{\rm ref},\mu_B)$ imposed at all orders. Hence, while both methods eventually converge to the correct values, we can say that the $s$-based method can reach a better convergence more quickly, as shown in Fig.~\ref{fig:nb_muconst}. In view of eventual applications to large-scale projects with more expensive lattices, we note that it is extremely valuable that the expansion to order $\calO (\mu_B^4)$ agrees almost perfectly with the $\calO (\mu_B^8)$ order, while coming at a fraction of the cost.

\section{Generalized $BQS$ directions}\label{sec:BQS}

\begin{figure*}
    \centering
    \includegraphics[width=0.49\linewidth]{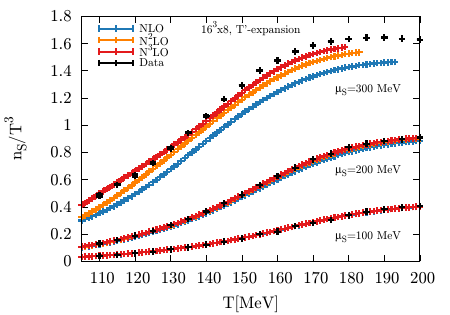}
        \includegraphics[width=0.49\linewidth]{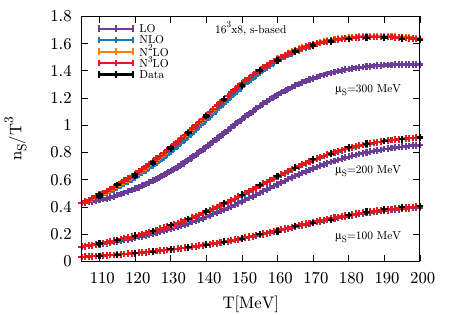}
    \caption{Strangeness density as a function of the temperature at constant strangeness chemical potentials (and $\mu_B=0$) from the $T^\prime$ (left) and $s$-based (right) expansions. Different colors correspond to different orders in the expansion, including up to $\chi_2^S$ (purple), $\chi_4^S$ (blue), $\chi_6^S$ (orange) or $\chi_8^S$ (red). Black points are obtained through the direct method of Ref.~\cite{Adam:2026rnt}.}
    \label{fig:ns_muconst}
\end{figure*}

So far we have considered the $T^\prime$ and $s$-based expansions varying the baryon chemical potential only, but both can easily be generalized to the case of three chemical potentials $\mu_B,\mu_Q,\mu_S$. The $T^\prime$ expansion was extended to this general case in Ref.~\cite{Abuali:2025tbd}, whereby the expansion was defined along an arbitrary direction in the 3D space of chemical potentials, parametrized in spherical coordinates by the angles $\theta,\varphi$. Susceptibilities $\chi_n^B$ are replaced by ``generalized'' susceptibilities $X^{\theta,\varphi}_n$ defined through derivatives with respect to the ``generalized'' chemical potential defined by the angles $\theta,\varphi$:
\begin{equation}
    X_n^{\theta,\varphi} (T) = \frac{\partial^n p/T^4}{\partial \hmu^n} \Bigg|_{\hmu=0}^{\theta,\varphi} \, \, ,
\end{equation}
with
\begin{align} \nonumber     \label{eq:mustoangles}
    \hat{\mu}_B &= \hat{\mu} \cos\theta \, \, ,\\
    \hat{\mu}_Q &= \hat{\mu} \sin \theta \cos \varphi \, \, , \\ \nonumber
    \hat{\mu}_S &= \hat{\mu} \sin \theta \sin \varphi \, \, ,
\end{align}
where the derivatives are taken at fixed values of $\theta,\varphi$. By replacing every instance of $\chi_n^B$ with the corresponding $X^{\theta,\varphi}_n$ all equations defining the expansion remain identical (see Ref.~\cite{Abuali:2025tbd} for details). This means that the procedure can (must) be repeated for every angle pair $\theta,\varphi$ of interest, redefining the generalized chemical potential and susceptibilities. The special case $n=1$ corresponds to the generalized charge density $\hat n_{\theta,\varphi} = X_1^{\theta,\varphi}$, and enters the defining equation of the scheme in a general direction:
\begin{align}
    \frac{X_1^{\theta,\varphi}(T, \hmu)}{\hmu} =
    X_2^{\theta,\varphi}(T^{\, \prime \, \theta,\varphi}(T,\hat{\mu}), 0)
    \, \, .
    \label{eq:4D-TExS}
\end{align}

Here emerges another substantial difference between the $T^\prime$ and $s$-based
expansion methods. Being based on a charge density, i.e. a derivative of the
free energy wrt a chemical potential, in the former method the expansion of
$T^\prime(T,\mu(\theta,\varphi))$ must be based on the same chemical potential
$\mu(\theta,\varphi)$ that defines the charge density. In other words, in
Eq.~\eqref{eq:Tprime} the baryon density normalized by the baryon chemical
potential forces the rhs to contain the second baryon susceptibility and
$T^\prime$ to contain an expansion in $\mu_B$, too. This is the reason why the
whole construction is rotated in 3D space in the procedure of
Ref.~\cite{Abuali:2025tbd}, i.e. the angles $\theta,\varphi$ in
Eq.~\eqref{eq:4D-TExS} are the same for $X_1,X_2$ and $T^\prime$. For the
$s$-based method this is different: not based on a charge density,
Eq.~\ref{eq:sbased_def} has no preferred direction built in, and
Eq.~\eqref{eq:sbased_Ts} can be directly written as a multi-dimensional
expansion as:
\begin{equation} \label{eq:sbased_Ts_BQS}
    T(T_0,\mu_B,\mu_Q,\mu_S) = T_0 + \sideset{}{^{\prime}}\sum_{n,m,l} \frac{\alpha_{nml} (T_0)}{n! \, m! \, l!} \mu_B^n \mu_Q^m \mu_S^l   \, \, ,
\end{equation}
where the primed sum is restricted to $(n+m+l)>1$ and even, due to charge
conjugation. A version of Eq.~(\ref{eq:sbased_Ts_BQS}) was
introduced in Ref.~\cite{Shah:2026mzu}, where the coefficients $\alpha_{nml}$ were defined as functions of $\theta$ and $\varphi$, rather than as in Eq.~(\ref{eq:sbased_Ts_BQS}). 
The above expression has the advantage that, if one is interested in mapping the
full 3D space of chemical potentials, the whole construction is used only once,
including the elaborate interpolation and fitting procedure described in Appendix~\ref{app:fit}.

In order to compare the convergence properties of both expansions we consider two distinct directions. First, we show the strangeness density as a function of the temperature, for different real strangeness chemical potentials and $\mu_B=0$ in Fig.~\ref{fig:ns_muconst}, which simply means replacing every instance of $B$ with $S$ in the construction of the previous Sections. In this case, the difference between the two schemes is much more pronounced, with the $s$-based method still achieving convergence very quickly, and the $T^\prime$ not doing so even at order $\calO (\mu_S^8)$.
As before, we also show the full grand canonical result obtained with the method of Ref.~\cite{Adam:2026rnt}. Up to $\mu_S\approx 300$~MeV we have perfect agreement for the case of the (already converged) $s$-based result, while for the $T^\prime$ expansion this is true only up to $\mu_S\approx 200$~MeV. We note that, in the $s$-based expansion, the ansatz for the integration constant $p_{\rm ref}$ we used in the previous Section must be replaced by a more general one including the strangeness chemical potential (see e.g. Ref~\cite{Alba:2017mqu}):
\begin{align}
    \hat{p}(T_{\rm ref},\mu_B) &= \hat{p}^{BS}_{00}(T_{\rm ref}) + \hat{p}^{BS}_{10}(T_{\rm ref}) \cosh(\mu_B/T_{\rm ref}) \\ \nonumber
    &\quad + \hat{p}^{BS}_{01}(T_{\rm ref}) \cosh(\mu_S/T_{\rm ref}) \\ \nonumber
    &\quad + \hat{p}^{BS}_{11}(T_{\rm ref}) \cosh(\mu_B/T_{\rm ref} - \mu_S/T_{\rm ref})\\ \nonumber
    &\quad + \hat{p}^{BS}_{12}(T_{\rm ref}) \cosh(\mu_B/T_{\rm ref} - 2\mu_S/T_{\rm ref})\\ \nonumber
    &\quad + \hat{p}^{BS}_{13}(T_{\rm ref}) \cosh(\mu_B/T_{\rm ref}-3\mu_S/T_{\rm ref}) 
\end{align}
with the partial pressures $\hat{p}^{BS}_{ij}$ fixed by the baryon and strangeness diagonal and off-diagonal susceptibilities, and $\hat{p}^{BS}_{00}$ fixed by the equation of state at $\mu_B=\mu_S=0$.

\begin{figure*}
    \centering
    \includegraphics[width=0.49\linewidth]{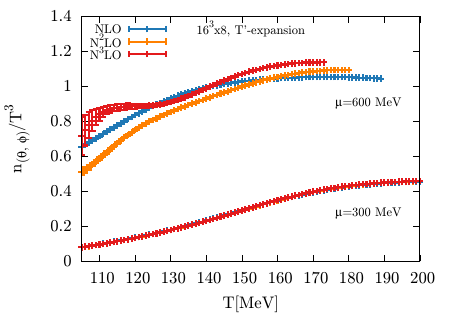}
        \includegraphics[width=0.49\linewidth]{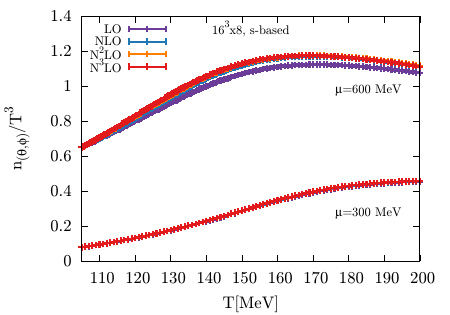}
    \caption{Generalized density as a function of the temperature at constant generalized chemical potential defined by $\theta=\pi/4,\varphi=\pi/2$ from the $T^\prime$ (left) and $s$-based (right) expansions. Different colors correspond to different orders in the expansion, including up to $\chi_2$ (purple), $\chi_4$ (blue), $\chi_6$ (orange) or $\chi_8$ (red).}
    \label{fig:nX_muconst}
\end{figure*}

Finally, in Fig.~\ref{fig:nX_muconst} we carry out an analogous
comparison in the direction defined by $\theta=\pi/4,\varphi=\pi/2$. We note that the convergence in the generalized density $\hat n_{\theta,\varphi}$ is still excellent for the $s$-based method, while for the $T^\prime$-expansion, although better than in the $S$ case, it is still substantially worse than in the case with $B$ only.

\section{Conclusions}\label{sec:conclusion}

In this work we have exploited our extreme-statistics dataset on a $\lat{16}{8}$ lattice to construct expansions for the QCD equation of state to order $\calO (\mu^8)$ in the chemical potential(s) and compared different prescriptions for the resulting extrapolations. Our main goals were to determine if at this fixed order the expansion schemes agree, and whether one would converge more quickly, already at a lower order in the expansion. This is especially important in view of possible large-scale applications on larger lattices, where the numerical cost is the true bottleneck. 

We compared the $T^\prime$ expansion of Ref.~\cite{Borsanyi:2021hbk} and the $s$-based expansion of Refs.~\cite{Shah:2024img,Shah:2026pue}, both in the case of a single chemical potential, as well as in the general case of multiple chemical potentials. Previously, the $T^\prime$ expansion had been formally derived and used up to N$^2$LO in Refs.~\cite{Borsanyi:2021hbk,Borsanyi:2022qlh}, and the $s$-based scheme had been worked out up to LO~\cite{Shah:2024img,Shah:2026pue}. In this work we extended both to N$^3$LO and presented data up to these high orders for the first time.

We showed that both methods are largely superior to the naive Taylor expansion, as was noted already in Ref.~\cite{Borsanyi:2021hbk} for the $T^\prime$ expansion.  We found that the two methods agree at the highest order $\calO (\mu^8)$ in the case of baryon chemical potential at least up to $\mu_B\approx 550$~MeV, but the $s$-based expansion reaches convergence more quickly, already at order $\calO (\mu^4)$. This rapid convergence persists both in the case of finite $\mu_S$ and $\mu_B=0$, as well as in a ``mixed'' direction in the $\mu_B-\mu_S$ plane defined by the angles $\theta=\pi/4,\varphi=\pi/2$, while in both directions a perfect agreement between the two methods could not be reached even at the highest order. Wherever full, extrapolation-free grand canonical results from Ref.~\cite{Adam:2026rnt} were available, we cross-checked the expansions against them. The $T^\prime$ expansion was in agreement with this benchmark if it converged. The highest order of the $s$-based expansion was always in agreement in the studied range, but even the next-to-leading-order provided an extremely good approximation.

From this evidence we can conclude that the method based on entropy contours generally offers a more rapid convergence. Adding only one order to the original proposal in Ref.~\cite{Shah:2024img} one obtains an expansion scheme with a range of applicability up to $\mu_B\approx$~500~MeV. As input, the expansion to this order requires the knowledge of quark number susceptibilities up to fourth order (see~\cite{Bellwied:2015lba,Bollweg:2021vqf,Bollweg:2022rps,Abuali:2025tbd,Borsanyi:2026jbj}) and the equation of state at $\mu=0$ \cite{Borsanyi:2025dyp}. While the knowledge of higher order susceptibilities could extend this range, we showed that at least two additional orders are needed as input in order to significantly improve the accuracy. This extension, however, may increase the numerical cost by several orders of magnitude.

An additional appealing feature of the $s$-based expansion we observed, is its more agnostic nature with respect to the specific direction in the space of chemical potentials. These features ultimately mean that this approach is well suited to generalizations aimed at mapping the equation of state for arbitrary conditions.

\begin{acknowledgments}
Z. Fodor acknowledges funding from the DOE under the contract number DE-SC0025025.
This work was also supported by the
Hungarian National Research, Development and Innovation
Office, NKFIH Grant No. KKP126769.
This work was also supported by the NKFIH excellence
grant TKP2021{\textunderscore}NKTA{\textunderscore}64.
This work is also supported by the Hungarian National Research,
Development and Innovation
Office under Project No. FK 147164. A. P. acknowledges support of the Momentum Programme of the Hungarian
Academy of Sciences (grant LP2025-13/2025).
The authors gratefully acknowledge the Gauss Centre for
Supercomputing e.V. (\url{www.gauss-centre.eu}) for funding
this project by providing computing time on the GCS Supercomputer
Juwels-Booster at Juelich Supercomputer Centre.
We acknowledge the EuroHPC Joint Undertaking for awarding this project access to the EuroHPC supercomputer LUMI, hosted by CSC (Finland) and the LUMI consortium through a EuroHPC Extreme Access call.
An award of computer time was provided by the INCITE program.
This research used resources of the Argonne Leadership Computing Facility, which
is a DOE Office of Science User Facility supported under Contract DE-AC02-06CH11357.

\end{acknowledgments}


\appendix

\section{Alternative choices for \textit{implicit} and \textit{explicit} schemes} \label{app:schemes}

As we mentioned in the text, expansion schemes based on following contours can differ by whether the $\mu_B=0$ or the finite-$\mu_B$ temperatures are expanded, and we referred to these choices as \textit{explicit} and \textit{implicit}. We also noted that, historically, the $T^\prime$ expansion was introduced in the \textit{explicit} form, and the $s$-based in the \textit{implicit} one. In this Section we explore the opposite choices in terms of convergence.

\begin{figure*}
    \centering
    \includegraphics[width=0.45\linewidth]{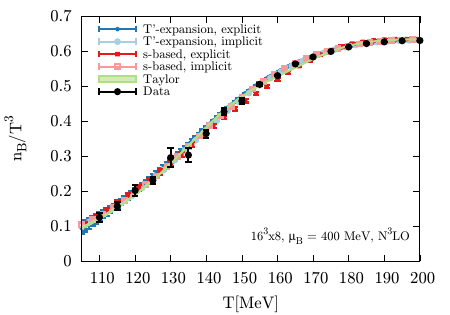}
    \includegraphics[width=0.45\linewidth]{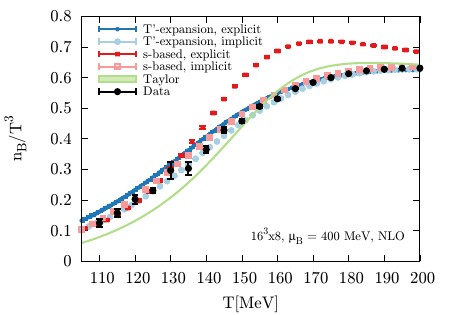}
    \includegraphics[width=0.45\linewidth]{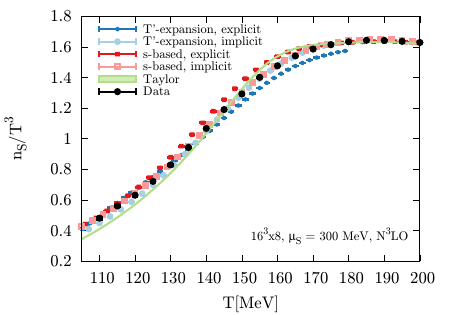}
    \includegraphics[width=0.45\linewidth]{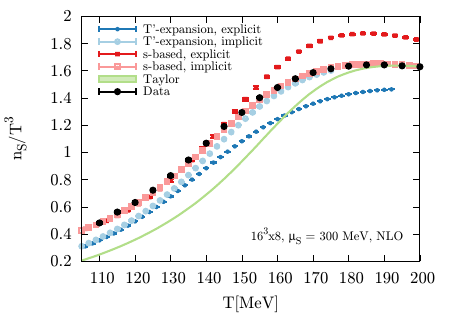}
    \caption{Baryon density (top panels) and strangeness density (bottom panels) as functions of the temperature at fixed $\mu_B=400$~MeV (top) or $\mu_S=300$~MeV (bottom), from the \textit{implicit} $T^\prime$ (light blue) and $s$-based (light red) expansions, as well as from the \textit{explicit} $T^\prime$ (dark blue) and $s$-based (dark red) expansions together with the Taylor result (light green). The black points are obtained through an extrapolation-free method based on a canonical approach in Ref.~\cite{Adam:2026rnt}. 
    The left panels compare the expansion schemes at NNNLO (using coefficients up to $\chi^{B/S}_8(T)$), the right panels are at NLO (using coefficients up to $\chi^{B/S}_4(T)$).
    \label{fig:imp_vs_exp}
    }
\end{figure*}

\subsection{Implicit scheme based on $n_B/\hmu_B$}

The defining equation for this scheme is:
\begin{equation}
    \frac{\chi_1^B (T^*,\hmu_B)}{\hmu_B} = \chi_2^B(T,0)
    \label{eq:nb_imp_matching}
\end{equation}
with the following expansion for $T^*$:
\begin{equation}\label{eq:Tstar}
    T^* (T,\hmu_B) = T \left( 1 + \sum_{n=2}^{\rm even} \tilde{\kappa}_n(T)\left(\frac{\mu_B}{T}\right)^n \right) \, \, .
\end{equation}
The temperature that normalizes $\mu_B$ in $\hmu_B$ is ambiguous. We have 
to use $\hmu_B=\mu_B/T^*$ in the expansion
\begin{equation}
    \frac{\chi_1^B (T^*,\hmu_B)}{\hmu_B} = \sum_{n=2}^{\rm even} \chi_{n+2}^B(T^*)\frac{\hmu_B^n}{(n+1)!} \, ,
    \label{eq:nb_imp_chi_expand}
\end{equation}
however, in Eq.~(\ref{eq:Tstar}) we deliberately expand in $\mu_B/T$ (not $\hat \mu_B$).
We have checked numerically, that an expansion of $T^*$ in $\mu_B/T^*$ would result in
much inferior convergence. We now expand each of the terms in Eq.~(\ref{eq:nb_imp_chi_expand})
in $\mu_B$, and matching the terms we obtain:
\begin{align}
    \chi_2^B(T) &= \chi_2^B(T)\\
    \chi_4^B(T) &= -6T\tilde{\kappa}_2 \chi_2^{B'}(T)\nonumber\\
    \chi_6^B(T) &= -120T \tilde{\kappa}_4 \chi_2^{B'}\nonumber \\
    &- 60T^2\tilde{\kappa}_2^2\chi_2^{B''} - 20T \tilde{\kappa}_2 \chi_4^{B'} + 40 \tilde{\kappa}_2 \chi_4^B\nonumber\\
    \chi_8^B(T) &= -5040 \left[ T \tilde{\kappa}_6 \chi_2^{B'} + T^2 \tilde{\kappa}_2 \tilde{\kappa}_4 \chi_2^{B''} + T^3 \tilde{\kappa}_2^3 \frac{\chi_2^{B'''}}{6} \right.\nonumber\\
    &- \tilde{\kappa}_2 \frac{\chi_6^B}{30} + T \tilde{\kappa}_2 \frac{\chi_6^{B'}}{120} + \tilde{\kappa}_2^2 \frac{\chi_4^B}{2} \nonumber \\
    &\left. - \tilde{\kappa}_4 \frac{\chi_4^B}{3} - T\tilde{\kappa}_2^2 \frac{\chi_4^{B'}}{3} + T \tilde{\kappa}_4 \frac{\chi_4^{B'}}{6} + T^2 \tilde{\kappa}_2^2 \frac{\chi_4^{B''}}{12}\right]\nonumber
\end{align}

After inserting the derivatives of $\chi^B_4$ and $\chi^B_6$ we have
\begin{align}
    \chi_2^B(T) &= \chi_2^B(T)\\
    \chi_4^B(T) &= -6T\tilde{\kappa}_2 \chi_2^{B'}\nonumber\\
    \chi_6^B(T) &= 60T^2 \tilde{\kappa}_2^2 \chi_2^{B''} \nonumber\\
    &+120T^2\tilde{\kappa}_2 \tilde{\kappa}_2' \chi_2^{B'} - 120T \tilde{\kappa}_4 \chi_2^{B'} - 120 T \tilde{\kappa}_2^2 \chi_2^{B'}\nonumber\\
    \chi_8^B(T) &= -840 T^3 \tilde{\kappa}_2^3 \chi_2^{B'''} + 5040 (T^2\tilde{\kappa}_2^3 + T^2 \tilde{\kappa}_2 \tilde{\kappa}_4 - T^3 \tilde{\kappa}_2^2 \tilde{\kappa}_2') \chi_2^{B''} \nonumber\\
    & +(-5040 T \tilde{\kappa}_6 + 15120 T^2 \tilde{\kappa}_2^2 \tilde{\kappa}_2'\nonumber\\
    &\,\,\, - 20160 T \tilde{\kappa}_2 \tilde{\kappa}_4 - 10080 T \tilde{\kappa}_2^3 - 5040 T^3 \tilde{\kappa}_2 \tilde{\kappa}_2'^2 \nonumber\\
    &- 2520 T^3 \tilde{\kappa}_2^2 \tilde{\kappa}_2'' + 5040 T^2 \tilde{\kappa}_2 \tilde{\kappa}_4' + 5040 T^2 \tilde{\kappa}_2' \tilde{\kappa}_4) \chi_2^{B'}\nonumber
\end{align}

Note that these expressions for $\chi_2^B$, $\chi_4^B$, $\chi_6^B$, $\chi_8^B$ are different from the corresponding ones in the $T^\prime$ expansion, Eq.~\ref{eq:chistokappas}, as they define a different expansion.

\subsection{Explicit scheme based on entropy density}

In this case the defining equation is still Eq.~\eqref{eq:sbased_def}:
\begin{equation} \label{eq:sbased_exp_def}
    s(T,\mu_B) = s(T_0,0) \, \, , 
\end{equation}
but now we expand $T_0$ as:
\begin{equation} \label{eq:sbased_T0}
    T_0 (T,\mu_B) = T + \sum_{n=2}^{\rm even} \frac{\tilde{\alpha}_n(T)}{n!} \mu_B^n \, \, .
\end{equation}

Now we match $s(T,\mu_B) = s(T_0, 0)$ up to order $\mathcal{O}(\mu_B^8)$. writing the dimensionful entropy at finite chemical potential as:
\begin{equation}
    s(T,\mu_B) = \sum_{n=0}^4 \mathcal{S}_{2n}(T)\mu_B^{2n} \, \, .
\end{equation}

Then, we start from the condition:
\begin{equation}
    \mathcal{S}_0(T) + \mathcal{S}_2(T)\mu_B^2 + \mathcal{S}_4(T)\mu_B^4 + \mathcal{S}_6(T)\mu_B^6 + \mathcal{S}_8(T)\mu_B^8= \mathcal{S}_0(T_0) \, \, ,
\end{equation}
where the rhs expands to:
\begin{align} \nonumber
    \mathcal{S}_0(T_0) &= \mathcal{S}_0(T) + \mathcal{S}_0'(T)(T_0-T) + \mathcal{S}_0''(T) \frac{(T_0-T)^2}{2} \\
    &\quad+ \mathcal{S}'''_0(T) \frac{(T_0-T)^3}{6} + \mathcal{S}_0^{(4)}(T) \frac{(T_0 - T)^4}{24} \, \, .
\end{align}
which equals, up to $\mu_B^8$:
\begin{align}    
    \mathcal{S}_0(T_0) &= \mathcal{S}_0(T) \\ \nonumber
    &\, + \mathcal{S}_0'(T)\Biggl[ \frac{\tilde{\alpha}_2}{2} \mu_B^2 + \frac{\tilde{\alpha}_4}{24}\mu_B^4 + \frac{\tilde{\alpha}_6}{720}\mu_B^6 + \frac{\tilde{\alpha}_8}{40320}\mu_B^8 \Biggr]\\ \nonumber
    &\, + \frac{\mathcal{S}_0''(T)}{2}\Biggl[\frac{\tilde{\alpha}_2^2}{4}\mu_B^4 + \frac{\tilde{\alpha}_2 \tilde{\alpha}_4}{24}\mu_B^6 + \left( \frac{\tilde{\alpha}_2 \tilde{\alpha}_6}{720} + \frac{\tilde{\alpha}_4^2}{576} \right) \mu_B^8\Biggr]\\ \nonumber
    &\, + \frac{\mathcal{S}_0'''(T)}{6} \Biggl[ \frac{\tilde{\alpha}_2^3}{8} \mu_B^6 + \frac{\tilde{\alpha}_2^2 \tilde{\alpha}_4}{32} \mu_B^8\Biggr] \\ \nonumber
    &\, + \frac{\mathcal{S}^{(4)}(T)}{24} \Biggl[ \frac{\tilde{\alpha}_2^4}{16} \mu_B^8 \Biggr] \, \, .
\end{align}

Matching orders with the lhs yields:
\begin{align}
    \mathcal{S}_2(T) &= \frac{\tilde{\alpha}_2}{2} \mathcal{S}_0'(T)\\ \nonumber
    \mathcal{S}_4(T) &= \frac{\tilde{\alpha}_4}{24} \mathcal{S}_0'(T) + \frac{\tilde{\alpha}_2^2}{8} \mathcal{S}_0''(T)\\ \nonumber
    \mathcal{S}_6(T) &= \frac{\tilde{\alpha}_6}{720}\mathcal{S}_0'(T) + \frac{\tilde{\alpha}_2 \tilde{\alpha}_4}{48}\mathcal{S}_0''(T) + \frac{\tilde{\alpha}_2^3}{48}\mathcal{S}_0'''(T)\\ \nonumber
    \mathcal{S}_8(T) &= \frac{\tilde{\alpha}_8}{40320}\mathcal{S}_0'(T) + \frac{\tilde{\alpha}_2 \tilde{\alpha}_6}{1440} \mathcal{S}_0''(T) + \frac{\tilde{\alpha}_4^2}{1152} \mathcal{S}_0''(T) \\ \nonumber
    &\qquad + \frac{\tilde{\alpha}_2^2 \tilde{\alpha}_4}{192} \mathcal{S}_0'''(T) + \frac{\tilde{\alpha}_2^4}{384} \mathcal{S}_0^{(4)}(T) \, \, ,
\end{align}
or, in terms of the fluctuations:
\begin{align}
    \frac{T^2 \chi_2^B}{2} &= \int \frac{\tilde{\alpha}_2}{2} \mathcal{S}_0'(T) dT + C_2\\  \nonumber
    \frac{\chi_4^B}{24} &= \int \left(\frac{\tilde{\alpha}_4}{24} \mathcal{S}_0'(T) + \frac{\tilde{\alpha}_2^2}{8} \mathcal{S}_0''(T)\right) dT+ C_4\\ \nonumber
    \frac{\chi_6^B}{720T^2} &= \int \left(\frac{\tilde{\alpha}_6}{720}\mathcal{S}_0'(T) + \frac{\tilde{\alpha}_2 \tilde{\alpha}_4}{48}\mathcal{S}_0''(T) \right. \\ \nonumber
    &\quad\qquad \left.+ \frac{\tilde{\alpha}_2^3}{48}\mathcal{S}_0'''(T)\right) dT + C_6\\ \nonumber
    \frac{\chi_8^B}{40320T^4} &= \int \left(\frac{\tilde{\alpha}_8}{40320}\mathcal{S}_0'(T) + \frac{\tilde{\alpha}_2 \tilde{\alpha}_6}{1440} \mathcal{S}_0''(T) + \frac{\tilde{\alpha}_4^2}{1152} \mathcal{S}_0''(T) \right. \\  \nonumber
    &\quad + \left. \frac{\tilde{\alpha}_2^2 \tilde{\alpha}_4}{192} \mathcal{S}_0'''(T) + \frac{\tilde{\alpha}_2^4}{384} \mathcal{S}_0^{(4)}(T) \right)dT + C_8 \, \, .
\end{align}

These relations complete the definition of both schemes, because they allow to compute the corresponding expansion coefficients and reconstruct the thermodynamics from the defining equations Eqs.~\eqref{eq:nb_imp_matching},~\eqref{eq:Tstar},~\eqref{eq:sbased_exp_def},~\eqref{eq:sbased_T0}.

In order to gauge the convergence properties of these expansions, we compare
them to the choices in the main text in Fig.~\ref{fig:imp_vs_exp}, showing the
baryon and strangeness densities as functions of the temperature at fixed chemical potentials. We
also include the full grand canonical results obtained in Ref.~\cite{Adam:2026rnt}.
In the left panels we compare the predictions of the expansions in the main text for the baryon density at fixed $\mu_B$ (top) and the strangeness density at fixed $\mu_S$ (bottom), 
as well as the alternatives introduced in this Appendix. At these high order all schemes
give a satisfactory agreement with the full result for $n_B$, while for $n_S$ visible disagreement appears in the \textit{explicit} methods. How fast the various schemes
converge to the full result is different, though. As we show in the right panels,
having only access to $\chi^{B/S}_4(T)$ (i.e., computing only NLO), the explicit $s$-based
method performs much worse, while the \textit{implicit} $T'$ expansion seems to outperform, especially for $n_S$, its \textit{explicit} counterpart. The $s$-based \textit{implicit} scheme shows excellent agreement with the direct data in all cases.

\section{Fit procedure} \label{app:fit}
It is clear from the expressions of the expansion coefficients for the schemes discussed above that up to third order derivatives with respect to the temperature are needed for quantities such as $\chi_2^B(T)$, which are known to us only at discrete temperature points. To this end, we employ a Bayesian fitting procedure using splines of degree $m=5$ with free knots using Reversible Jump Markov Chain Monte Carlo (RJMCMC). In the following discussion, we assume that we have $n$ observations $(x_i, y_i,\sigma_i)$, and the goal is to determine the model function $f(x)$ which describes the observations. We use B-spline (the basis defined by the Cox-de Boor recursion formula) with the following knot vector for a spline of degree $m$:
\begin{equation}
    t = (\underbrace{a,a,a,a, \dots}_{m+1 ~\text{times}}, \xi_1, \xi_2, \xi_3, \dots, \xi_k, \underbrace{\dots,b,b,b,b}_{m+1 ~ \text{times}}) \, \, ,
\end{equation}
where $a$ and $b$ are the $x$-values corresponding to the endpoints. Note that the boundary knots are repeated $m + 1$ times, which corresponds to imposing \textit{no} continuity conditions at the endpoints. We use the notation $(k,\vec{\xi})$ to represent the knot configuration where $k$ refers to the number of internal knots and $\vec{\xi}$ denotes their locations. The total number of free parameters (i.e. the number of spline coefficients) in this case is $(k+m+1)$. We denote the $j$-th function in a B-spline basis by $b_j$ and the design matrix by $B_{k,\vec{\xi}}$, of size $n \times (k+m+1)$, whose $i,j$ component is $b_j(x_i)$. The model function $f$ can then be written as:
\begin{equation}
    f(x) = \sum_{j=1}^{k+m+1} \beta_j b_j(x) \, \, ,
\end{equation}
for the spline coefficients vector $\vec{\beta}$. For a Bayesian formulation of the model, priors on the unknown quantities $(\vec{\beta},k,\vec{\xi})$ must be specified. The prior decomposition can be written as follows:
\begin{equation}
    \pi(\vec{\beta},k,\vec{\xi}) = \pi_\beta(\vec{\beta}|k,\vec{\xi}) \pi_\xi(\vec{\xi}|k)\pi_k(k) \, \, ,
\end{equation}
which factorizes into the three following pieces.

\begin{itemize}
    \item Prior on the number of internal knots: we choose a Poisson prior given by:
    \begin{equation}
        \pi_k(k) = \frac{\lambda^k e^{-\lambda}}{k!} \, \, ,
    \end{equation}
    where $\lambda$ is the Poisson parameter that corresponds to the expected number of internal knots in order to describe the data. 
    \item Prior on the location of internal knots: suppose we have $n$ observations. We first divide the range of $x_i$ into $(n-1)$ intervals. Each interval can have at most one knot. For $k$ knots, we first pick $k$ intervals out of the $(n-3)$ valid intervals with uniform probability ($(n-1-2)$ valid intervals because we have the repeated knots at the $x_{min}$ and $x_{max}$, thus the first and last intervals do not have other knots). Inside the chosen interval $[x_i, x_{i+1}]$, the knot can be placed anywhere in the range $[x_{i}+\epsilon, x_{{i+1}}-\epsilon]$ with uniform probability. The tunable parameter $\epsilon>0$ ensures that there is a minimum separation of $2\epsilon$ between two knots.
    \item Prior on the B-spline coefficients: we use a unit-information prior on the coefficients, $\vec{\beta}$, given by:
    \begin{align}
    \pi_\beta (\vec{\beta} | k, \vec{\xi}) &= \mathcal{N}(0, n(B^T C^{-1} B)^{-1}) \nonumber \\
    &= \frac{\exp \left( -\frac{1}{2} \vec{\beta}^T \left( \frac{B^T C^{-1} B}{n} \right) \vec{\beta} \right)}{\sqrt{\det (2\pi n (B^T C^{-1} B)^{-1})}}
    \end{align}
    where $B$ is the design matrix (with suppressed knot configuration index) and $C$ is the covariance matrix. This choice of prior allows us to marginalize the full posterior distribution over the spline coefficients:
    \begin{equation}
    p(y|k,\vec{\xi}) = \int p(y|\vec{\beta},k,\vec{\xi}) \pi_\beta(\vec{\beta}|k,\vec{\xi}) d\vec{\beta} \, \, ,
    \end{equation}
    to obtain:
    \begin{align}
    p (y|k,\vec{\xi}) &= \frac{(n+1)^{-(k+m+1)/2}}{\sqrt{\det (2\pi C)}} \exp \left(  - \frac{1}{2} y^T \{ C^{-1} \right. \nonumber \\ 
    &\left. - \frac{n}{n+1} C^{-1} B (B^T C^{-1} B)^{-1} B^T C^{-1} \} y\right) \, \, .
    \end{align}

    Using the large-$n$ approximation $n/(n+1) \approx 1$, the expression takes the familiar form:
    \begin{equation}
        p(y|k,\vec{\xi}) \propto \exp\left(-\frac{1}{2} \text{BIC}(k,\vec{\xi})\right)
    \end{equation}
    where $\text{BIC}(k, \vec{\xi})$ is the Bayesian Information Criterion for the knot configuration $(k,\vec{\xi})$.
\end{itemize}

The next part is to assign the probabilities of a \textit{birth}, \textit{death} and \textit{move} step, i.e. of adding, removing or moving a knot. More concretely, given the current configuration with $k$ internal knots, what is the probability of doing a birth, death or a move step? We follow \cite{10.1093/biomet/88.4.1055} and have proposal probabilities for each of these steps given by:
\begin{align}
    b_k &= c ~\min \{ 1, \pi_k(k+1)/\pi_k(k)\} \nonumber \, \, , \\
    d_k &= c ~ \min \{ 1, \pi_k(k-1)/\pi_k(k)\} \nonumber \, \, , \\
    \eta_k &= 1 - d_k - b_k \, \, .
\end{align}

These are chosen to satisfy $b_k \pi_k(k) = d_{k+1} \pi_k(k+1)$. With a Poisson prior on the number of knots, these probabilities take the simple form:
\begin{align*}
    b_k &= c \min \{1,\lambda/(k+1)\}\nonumber \, \, , \\
    d_k &= c \min \{1,k/\lambda\} \nonumber \, \, , \\
    \eta_k &= 1-d_k-b_k \, \, ,
\end{align*}
with $c=0.4$ (chosen such that $2c < 1$). In a birth step, a new knot is placed in one of the currently empty valid intervals, chosen with uniform probability, at a uniformly distributed position within that interval. A death step removes one of the $k$ existing knots, chosen with uniform probability, and a move step relocates one knot within the same interval. The acceptance probability can then be decomposed into various parts as follows:
\begin{align} \nonumber
    \alpha &= \min \{ 1, (\text{likelihood ratio}) \times (\text{prior ratio}) \times \\ 
    &\qquad\qquad\qquad \times (\text{proposal ratio}) \} \, \, .
\end{align}

For our choice of priors and the probabilities for various steps, the acceptance probability simplifies to:
\begin{equation}
    \alpha = \min \{ 1, (\text{likelihood ratio}) \} = \min \{ 1, \exp (- \Delta \text{BIC}/2) \}
\end{equation}

We run the program on the central sample of our jackknife datasets to generate a sequence of $N_{\text{conf}}$ knot configurations labeled by $\{ (k_i,\vec{\xi}_i)\}$. For each knot configuration, we refit all the jackknife samples to get the mean and the jackknife standard deviation $\{ m_i, \sigma_i\}$ for any quantity $y$ derived from the fit (e.g. $f(x)$ or its derivatives at a given $x$). We then define a joint probability distribution for $y$ as follows:
\begin{equation}
    p(y) = \sum_i w_i ~\mathcal{N}(y; m_i, \sigma_i) \, \, ,
\end{equation}
where $w_i = 1/N_{\text{conf}}$ are the weights of various knot configurations. The Cumulative Distribution Function (CDF) then reads:
\begin{equation}
    P(y) = \int_{-\infty}^y dy' \sum_i w_i~ \mathcal{N}(y';m_i,\sigma_i) \, \, .
\end{equation}

Finally, we take the median of the CDF as the central value for $y$, and the combined error $\sigma$ from:
\begin{equation}
    \sigma_{\text{tot}}^2 = [\frac{1}{2} (y_{84} - y_{16})]^2 \, \, ,
\end{equation}
with $P(y_{16}) = 0.16$ and $P(y_{84}) = 0.84$.

\begin{figure*}
    \centering
    \includegraphics[width=0.49\linewidth]{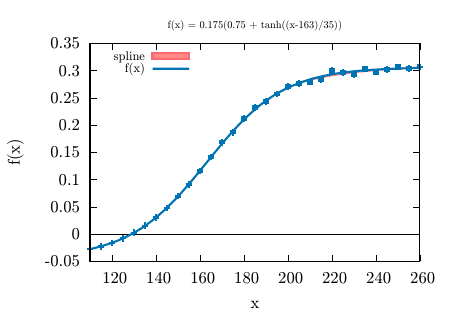}
        \includegraphics[width=0.49\linewidth]{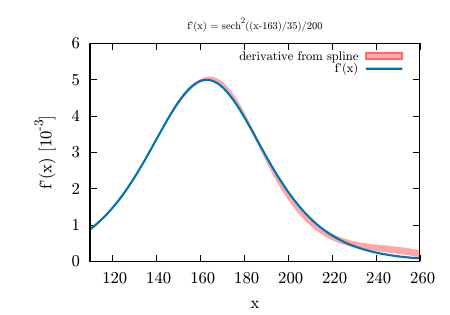} \\
            \includegraphics[width=0.49\linewidth]{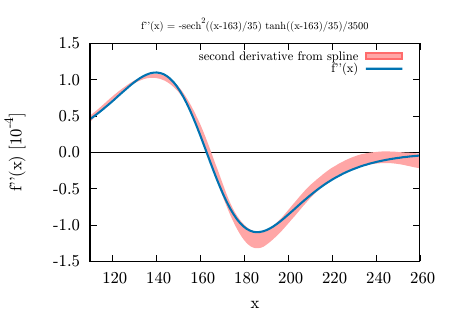}
        \includegraphics[width=0.49\linewidth]{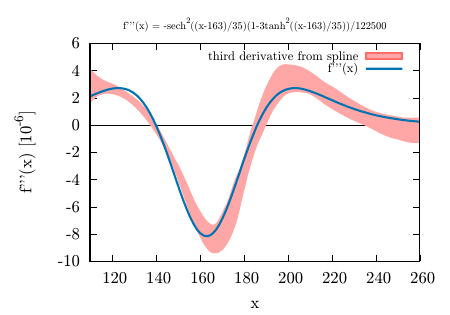}
    \caption{
    Mock data analysis showing the performance of the Bayesian fitting
    of quintic splines. The input data is based on a known sigmoid function with
    added noise (top left). This is fitted with error without assuming any
    specific functional form. The result is an ensemble of splines. The other
    panels show the first, second and third derivatives of the fitted splines
    as red bands. The width of these bands also reflect the variation of the splines,
    within which all known functional forms (blue lines) are reproduced.
    }
    \label{fig:sigmoid_test}
\end{figure*}

This concludes the discussion for the case when we have observations for a quantity $y$ and we wish to interpolate using splines with free knots. We show in Fig. \ref{fig:sigmoid_test}an application of this method on a mock data set for a sigmoid function. For our present use case, we first express the observables that we measure on the lattice in terms of the coefficients -- $\kappa_n (T)$ or $\alpha_n(T_0)$ -- depending on the scheme. As an illustration, we summarize the steps of this method in the case of the T'-expansion, for which the expressions are given by Eq. \eqref{eq:chistokappas}. Starting with the spline form for the coefficients, we have the following steps at each RJMCMC sweep:
\begin{itemize}
\item Systematically pick the spline object to modify.
\item Propose a birth, death or move step with the above mentioned probabilities and modify the knot configuration. Given the new knot configuration, minimize the joint chi-square given by $\chi^2 = \sum_{i=1}^{N_T} r_i^T \Sigma_i^{-1} r_i$ where $r_i = (r_{(2,i)}, r_{(4,i)}, r_{(6,i)}, r_{(8,i)})^T$ is the residual column vector at temperature $T_i$ and $\Sigma_i$ is a $4 \times 4$ covariance matrix which encodes the correlation between the susceptibilities which come from the same gauge configurations.
\item Compute the BIC for the knot configuration, followed by the accept-reject step. 

\end{itemize}

\bibliographystyle{unsrt}
\bibliography{thermo,paolo}

\end{document}